\documentclass[preprint,amsmath,amssymb,superscriptaddress,aps,floatfix]{revtex4-1}
\usepackage{xr} % For supplement materials
\usepackage{rotating}
\usepackage[dvipsnames]{xcolor}
\usepackage[title]{appendix}
\usepackage{ragged2e}
\usepackage{float}
\usepackage{chemformula}[2014/04/08]
\usepackage{subfigure}
\usepackage{tabularx}
    \newcolumntype{L}{>{\raggedright\arraybackslash}X}
    \newcolumntype{C}{>{\centering\arraybackslash}X}
    \newcolumntype{R}{>{\raggedleft\arraybackslash}X}
    \usepackage{siunitx}

\usepackage{dcolumn}% Align Table columns on decimal point
\usepackage{bm}% bold math
\usepackage[]{color}
\graphicspath{ {.//} }
\usepackage[utf8]{inputenc}
\definecolor{ultramarine}{RGB}{0,32,96}

\newcommand{\beginsupplement}{%
        \setcounter{table}{0}
        \renewcommand{\thetable}{S\arabic{table}}%
        \setcounter{figure}{0}
        \renewcommand{\thefigure}{S\arabic{figure}}%
     }

\begin{document}
%%%%%%%%%%%%%%%%\tracingall
\setchemformula{kroeger-vink}

\title{Optimizing ionic conductivity of lithium in Li$_7$PS$_6$ argyrodite via dopant engineering
}
% Force line breaks with \\

%\affiliation{
%Theory and Simulations of Materials %(THEOS), EPFL, Lausanne, Switzerland %\\
%}
%\affiliation{IBM RSM Zurich Research %Laboratory, Zurich, Switzerland}

\author{Sokseiha Muy}
\affiliation{
Theory and Simulations of Materials (THEOS) and National Centre for Computational Design and Discovery of Novel Materials (MARVEL), École Polytechnique Fédérale de Lausanne, CH-1015 Lausanne, Switzerland}%
%\altaffiliation[Also at ]{Physics Department, XYZ University.}%Lines break
\author{Thierry Le Mercier}
\affiliation{
Syensqo R\&I, 52 rue de la Haie Coq, 93300 Aubervilliers , France}
\author{Marion Dufour}
\affiliation{
Syensqo R\&I, 52 rue de la Haie Coq, 93300 Aubervilliers , France}
\author{Marc-David Braida}
\affiliation{
Syensqo R\&I, 52 rue de la Haie Coq, 93300 Aubervilliers , France}
\author{Antoine A. Emery}
\affiliation{
Syensqo R\&I, Rue de Ransbeek 310, 1120 Brussels, Belgium}
\author{Nicola Marzari}
\affiliation{
Theory and Simulations of Materials (THEOS) and National Centre for Computational Design and Discovery of Novel Materials (MARVEL), École Polytechnique Fédérale de Lausanne, CH-1015 Lausanne, Switzerland} %

\date{\today}% It is always \today, today,
\vspace{10cm}
\clearpage
\newpage
\begin{abstract}
Li-containing argyrodites represent a promising family of Li-ion conductors with several derived compounds exhibiting room-temperature ionic conductivity $>$ 1 mS/cm and making them attractive as potential candidates as electrolytes in solid-state Li-ion batteries. Starting from the parent phase Li$_7$PS$_6$, several cation and anion substitution strategies have been attempted to increase the conductivity of Li ions. Nonetheless, a detailed understanding of the thermodynamics of native defects and doping of Li argyrodite and their effect on the ionic conductivity of Li is missing. Here, we report a comprehensive computational study of defect chemistry of the parent phase Li$_7$PS$_6$ in both intrinsic and extrinsic regimes, using a newly developed workflow to automate the computations of several defect formation energies in a thermodynamically consistent framework. Our findings agree with known experimental findings, rule out several unfavorable aliovalent dopants,  narrowing down the potential promising candidates that can be tested experimentally. We also find that cation-anion co-doping can provide a powerful strategy to further optimize the composition of argyrodite. In particular, Si-F co-doping is predicted to be thermodynamically favorable; this could lead to the synthesis of the first F-doped Li-containing argyrodite. Finally, using DeePMD neural networks, we have mapped the ionic conductivity landscape as function of the concentration of the most promising cation and anion dopants identified from the defect calculations, and identified the most promising region in the compositional space with high Li conductivity that can be explored experimentally.
\end{abstract}

% \pacs{Valid PACS appear here}% PACS, the Physics and Astronomy

\maketitle

\section{\label{Intro}Introduction}
A global decarbonization of the world economy requires a massive transition to renewable energy sources, such as solar and wind as well as an electrification of the world transportation sectors, which in turn requires reliable and affordable technologies to store electricity. Conventional lithium-ion batteries using liquid electrolytes are a huge technological and commercial success; nevertheless, their performance rapidly approaching the theoretical limit and the use of highly flammable organic solvents as liquid electrolytes poses significant safety risks \cite{Janek}. One promising technology that has been actively researched and developed for the next-generation electricity-storage devices is that of all-solid-state batteries, with inorganic solid ion conductors which are intrinsically non flammable and have superior thermal stability. These would provide improved safety and better device integration by eliminating bulky components and the packaging required to cool the battery and to avoid leakage of liquid electrolytes. Moreover, the use of solid-state Li-ion conductors is believed to enable the replacement of graphite anodes by Li-metal anodes, which will further increase the energy density of the batteries \cite{SSB_review, ChenC6MH00218H, ZHENG2018198}. 

\begin{figure}
\centering
\includegraphics[width=0.9\linewidth]{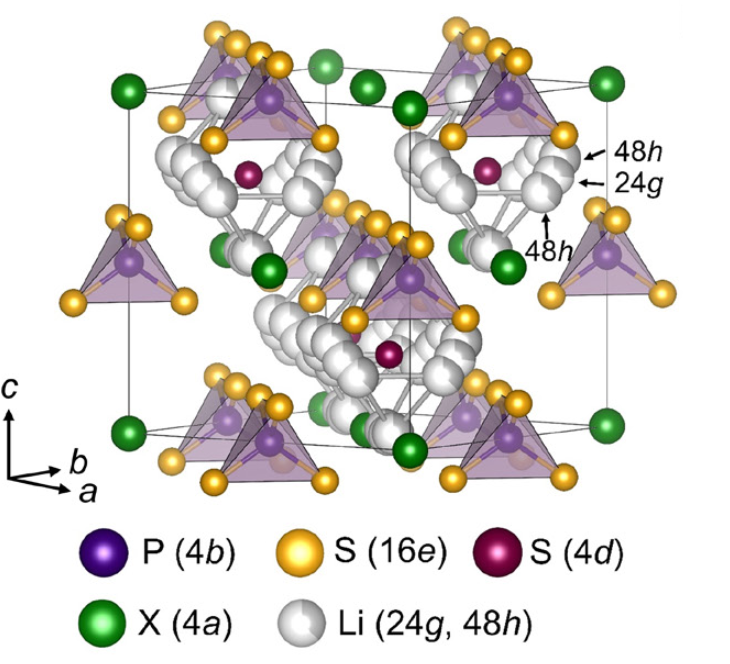}
% \vspace{1cm}
    \caption{Crystal structure of the cubic argyrodite of Li$_6$PS$_5$X, with the halide X sites forming a closed packed host lattice in which PS$^{3-}_4$ tetrahedra occupy the octahedral voids and half of the tetrahedral voids are filled with the free S$^{2-}$. When X = Cl or Br, there is a partial
    occupancy of the free S$^{2-}$ and X on the 4d and 4a sites which is believed to have significant influence on the long-range diffusion of Li ions \cite{Ohno_review}}. 
    \label{fig:crystal_structure}
\end{figure}

Several family of Li-containing compounds have been studied in the past decades as Li-ion conductors, with several compounds exhibiting room-temperature (RT) ionic conductivities comparable to that of liquid electrolytes \cite{muy_review, Ohno_review, YU2021105858}. One prominent family of Li-ion conductors in which several highly conducting compounds were discovered is the Li Argyrodite family, with the general composition Li$^+_{12-2m-y-x}$M$_m^{2+}$(Y$^{y+}$Ch$^{2-}_4$)Ch$^{2-}_{2-x}$X$^{x-}$ (M=Ca, Mg; Y=P, As, Ge, Si, Sn, Sb; Ch=O, S, Se; X=Cl, Br, I, BH$_4$) \cite{YU2021105858, Deiseroth, Ohno, Kraft, Kong_Li6PO5Cl, Kong_As, Kong_Li7PS6, Bernges, Sakuda, Zhou_Sb}. These can be derived from the parent compound Li$_7$PS$_6$ by substitution on the P-sites or S-sites or both \cite{Deiseroth_Li7PS6, Kong_Li7PS6}. The low-temperature phase of the parent compound (space group $Pna2_1$) has only a moderate RT ionic conductivity, on the order of $1.6 10^{-3}$ mS/cm \cite{Deiseroth_Li7PS6}. The high-temperature (HT) cubic phase (space group $F\bar{4}3m$) exhibits, on the other hand, high Li conductivity and can be stabilized at RT by partial substitution of S by the halide atoms (Cl, Br and I) \cite{Deiseroth, Kraft}. The (conventional) cubic unit cell then consists of a face-centered cubic (FCC) structure formed by the halides (Wyckoff position 4a) with MCh$^3_{4-}$tetrahedra (M on position 4b) occupying the octahedral voids as seen in figure \ref{fig:crystal_structure}. Half of the tetrahedral voids are filled by S$^{2-}$ (Wyckoff 4d, so called free sulfur sites) and the partially occupied lithium-ion positions (24g and 48h) form a cage around the free sulfur sites (see figure \ref{fig:crystal_structure}). Alternatively, the immobile host lattice can be described as a FCC lattice of PS$_4$ tetrahedra with the remaining free sulfur and halide ions occupying the octahedral and tetrahedral interstitials. The distribution of Li ions in Li$_6$MCh$_5$X can be well captured having only 48h sites occupied by 50\%; depending on the unit cell volume and material composition, occupancies on the 24g site can also be found \cite{Deiseroth, Kraft}. Within this cage-like isotropic compound, different ionic conduction pathways exist involving three possible jumps: 1) the doublet (48h–24g–48h), 2) the intra-cage (48h–48h), 3) and the inter-cage one (also 48h–48h).

To increase the ionic conductivity of the argyrodite, several doping strategies and synthesis routes have been developed to vary the concentration of Li ions as well as to control the distribution of sulfur and halide ions on the 4a and 4d sites. Despite all these efforts, a detailed understanding of the native defect chemistry as well as doping response in Li-Argyrodites is lacking. For instance, dopants are usually assumed to affect lithium stoichiometry by causing the formation of charge-compensating lithium vacancies or interstitials; for example, substitution of divalent anions S$^{2-}$ by monovalent anions Cl$^{-}$ is assumed to result in the creation of Li$^{+}$ such that the overall system remains charge-neutral. While this assumption seems to be verified under most synthesis conditions, there is no guarantee that it is the only doping response especially with new synthesis routes and thermal treatments which aim to stabilize/control a particular distribution of the sulfur and halide anions over the octahedral and tetrahedral sites in the argyrodite.

Moreover, simple charge compensation models usually assume that defects exist in their formal charge state for example, -1 for Li vacancies, or +2 for S vacancies which might be a good approximation in very ionic compounds such oxides and halides. Sulfides are much more covalent in nature, and these formal charges might not be a good approximation of the true defect charge states nor be consistent with the most stable charge states as a function of the Fermi level and the chemical potentials that reflect synthesis conditions. In fact, an accurate treatment of defect charges has to take into account a broad range of intrinsic and extrinsic defects in all accessible charge states whose concentration is determined self-consistently by the constraint of phase stability at the thermodynamic equilibrium and charge neutrality \cite{BUCKERIDGECHEMICALPOTENTIAL, BUCKERIDGEFERMI, FreysoldtRMP}. 

\section{\label{sec:section2} Methods}
\subsection{\label{experimental_section} Synthesis and Experimental Characterizations}
In an Ar-filled glovebox ($<$ 1 ppm H$_2$O, $<$1 ppm O$_2$), the precursors Li$_2$S (Lorad Chemical), LiCl (Sigma-Aldrich), and P$_2$S$_5$ (Sigma-Aldrich) were weighed and mixed in stoichiometric proportions to obtain 4g of the desired compositions. The mixture was then transferred into a 45 mL ZrO$_2$ jar filled with 5 mm zirconia (YSZ) balls, maintaining a ball-to-powder ratio of 16.5. The sealed jar was removed from the glovebox and placed in a Fritsch Planetary Micro Mill Pulverisette 7, where it was ball milled at 500 RPM for 2 hours, with 15-minute breaks every 30 minutes. The powder was recovered in the Ar-filled glovebox and the resulting powder was placed in a closed SiC crucible, heated to 500°C at a rate of 5°C/min in a tube furnace under N$_2$ atmosphere, and maintained at this temperature for 12 hours. The sample was cooled to RT, recovered in the Ar-filled glovebox, and deagglomerated in a mortar. 

X-ray diffraction of the samples was collected with an Aeris Research diffractometer from Malvern Panalytical (Cu, 600 W, 40 kV, 15 mA, Soller slit 0.02°). Diffractograms were collected in the 10° to 90° range over 2 hours. Lattice parameters were determined by fitting the diffraction profiles using FullProf Suite, with profile fitting performed using the F-43M space group of argyrodite. For impedance spectroscopy, the powders were sandwiched between two pre-dried carbon paper electrodes (Papyex soft graphite, 0.2 mm thick from Mersen) and cold-pressed into a 6 mm die at 500 MPa (Uniaxial manual press MS15-MDD, Eurolabo). The obtained pellets were loaded into an airtight measurement cell (EQ-PSC from MTI). All measurements were performed under a pressure of 40 MPa. AC impedance spectra were collected using a Biologic VMP3, with the sample temperature controlled by a Binder thermostatic chamber. The cell was connected to a galvanostat-potentiostat, and PEIS spectra were recorded with a 20 mV sinusoidal perturbation around OCV from 1 MHz to 10 kHz, with 25 points per decade, each point averaged from 50 measures. A model circuit was used to fit the curve and extract ionic resistance. Temperature tests were performed with 2-hour steps at each temperature, recording a spectrum at the end of each step. The temperature ranged from -20°C to 60°C, then back to 30°C. Electronic conductivity was measured by DC measurement at 2 volts, determined after 1 hour by the asymptotic method.

\subsection{\label{computation_section} Defect Calculations and Machine Learning Molecular Dynamics}
In order to compute defect concentrations, one needs to calculate the formation energy of those defects. The formalism of the defect formation energy (in dilute limit) is well established and can be obtained using the following relation \cite{Northrup_formalism, FreysoldtRMP}:
\begin{equation} \label{eq:FormationEnergy}
\begin{split}
E^f = E_{tot}[X_q]-E_{tot}[bulk]-\sum_in_i\mu_i +\\ 
    q(E_F+E_{vbm}+\Delta V_{pot})+E_{corr}^q
\end{split}
\end{equation}

where $E^f[X^q ]$ is the formation energy of defect X in charge state q, $E_{tot}[X^q ]$ is the (DFT) energy of the defected supercell, $E_{tot}[bulk]$ is the (DFT) energy of the pristing supercell without defect, $n_i$ are the number of atoms of type \emph{i} added ($n_i>0$) or removed ($n_i<0$) from the supercell, $\mu_i$ are the chemical potentials of atomic species \emph{i}, and $E_F$ is the Fermi level relative to the maximum of the valence band $E_{vbm}$. $\Delta V_{pot}$ and $E_{corr}^q$ are post-processing corrections which account respectively for the spurious interaction between the charge defect and the uniform background charge density, and the long-range electrostatic interactions between charge images in supercell calculations with periodic-boundary condition \cite{freysoldt_PRL_2009, freysoldt_2011, dabo_electrostatics_2008}. The chemical potential $\mu_i$ is the energy associated with exchanging atoms of species \emph{i} between the materials and a reservoir. Their values reflect the conditions under which the materials are synthesized and are constrained by the stability of the materials and that of all the competing phases as given by the phase diagram \cite{Northrup_formalism, BUCKERIDGECHEMICALPOTENTIAL}. 

To determine the Fermi level, we employ the fact that at thermodynamic equilibrium, in the presence of multiple defects with multiple charge states, its value is determined by requiring charge neutrality for the system \cite{BUCKERIDGEFERMI}
% as shown schematically in Figure \ref{fig:FermiLevel}. 
More precisely, if we denote $n_o$ the concentration of electrons, $p_o$ the concentration of holes and $C_{X^q}$ the concentration of defect X in the charge state q, the charge neutrality condition imposes that

\begin{equation} \label{eq:ChargeNeutrality}
n_o - \sum_X\sum_q qC_{X^q} = p_o
\end{equation}

where:
\begin{equation} \label{eq:c_electron}
n_o=\int_{E_F}^{\infty}f_e(E)\rho(E)dE
\end{equation}
\begin{equation} \label{eq:c_hole}
p_o=\int_{-\infty}^{E_F}f_h(E)\rho(E)dE
\end{equation}
\begin{equation} \label{eq:c_defect}
C_{X^q}=N_{site}\exp{\left(-\frac{E^f[X^q]}{kT}\right)}
\end{equation}

and $f_e$ is the Fermi-Dirac distribution, $f_e(E)=[1+\exp((E_F-E)/kT)]^{-1}$, $f_h=1-f_e$ and $\rho(E)$ the density of states per unit volume. Each term $n_o$, $p_o$ and $C_{X^q}$ depends on $E_F$, and therefore the condition \eqref{eq:ChargeNeutrality} constitutes a (non-linear) equation in $E_F$ which can be solved numerically to determine the value of Fermi level that corresponds to a neutral system. Implicit in this formalism is the so-called rigid band approximation, where it is assumed that the only response of the electronic structure to the presence of defects is the rigid shift of the Fermi level to adjust to the total number of electrons while the band structure of the materials remains unchanged.

For the present defect calculations, in the intrisic regime, we have considered all the non-equivalent lithium vacancies and interstitials V$_{\text{Li}}$ and i$_{\text{Li}}$, phoshorus vacancies V$_\text{P}$, sulfur vacancies on the octahedral (V$_{\text{S}_{\text{oct}}}$) and tetrahedral (V$_{\text{S}_{\text{tet}}}$) sites as well as sulfurs that are bonded to phosphorus atoms (V$_{\text{S}_{1-3}}$). In the extrinsic regime, we have considered mono- and divalent substitutions on Li sites (X$_\text{P}$, X = Na, Cu, Mg, Zn, and Ca), tri- and tetravalent substitutions on P sites (X$_\text{P}$, X = Al, Ga, In, Y, La, C, Si, Ge, Sn, Y, La, Hf, Ce) and anion substitution on S sites (X$_\text{S}$, X = F, Cl, Br, I, O, N). Finally, we have also considered co-doping scenarios where 2 dopants are simultaneously incorporated into the parent phase Li$_7$PS$_6$. 

\begin{figure}
\centering
\includegraphics[width=0.95\linewidth]{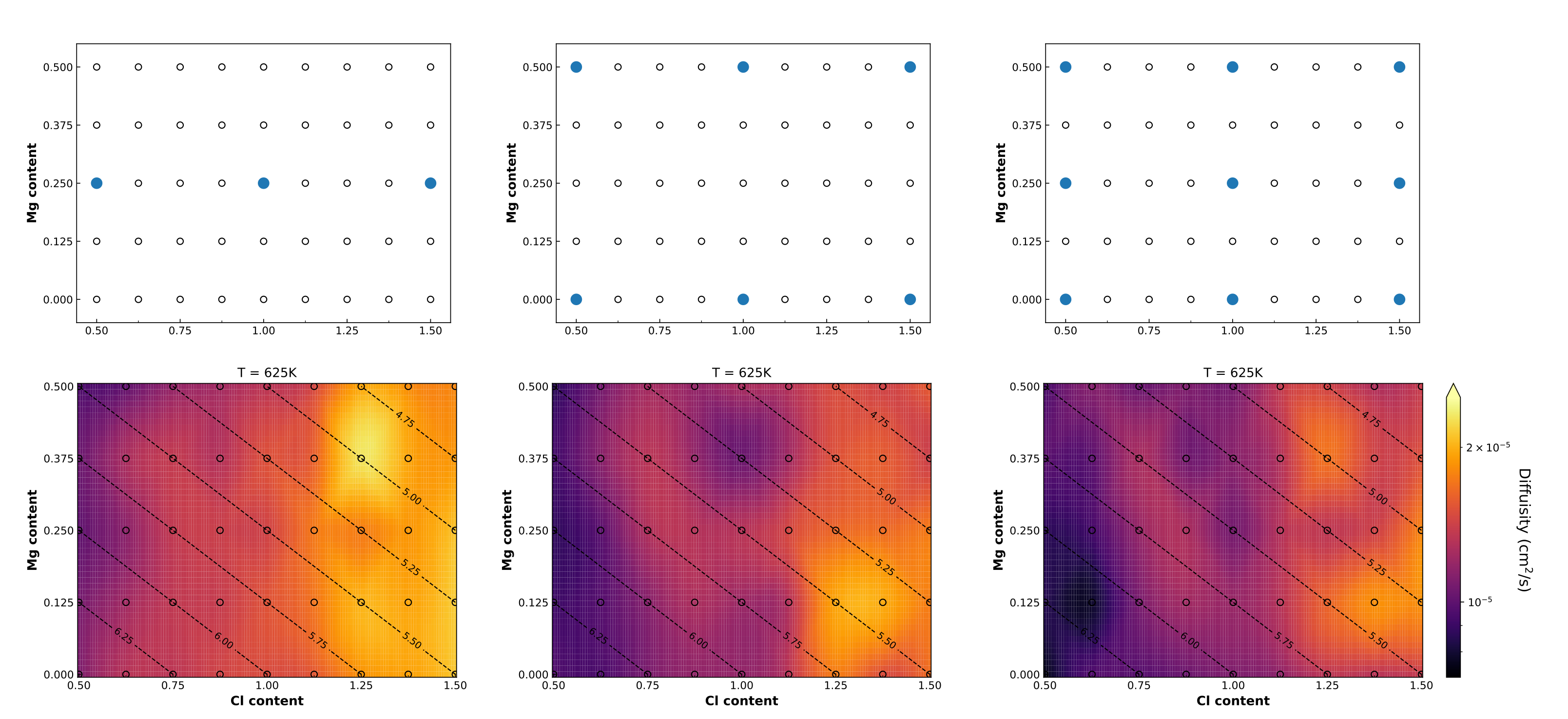}
    \caption{Ionic conductivity landscape obtained from 3 DeePMD potentials which were trained on 3, 6 and 9 compositions respectively (shown as the blue dots in the upper panels).}. 
    \label{fig:Cl_occupancy}
\end{figure}

To compute the ionic conductivity of lithium as a function of dopant concentrations, we used DeePMD \cite{DeePMD_kit, DeePMD_SE} a class of neural network potentials which have been used successfully to simulate the dynamics of Li-ions in the LGPS family \cite{DeePMD_LGPS} as well as other superionic conductors such as Na$_3$PS$_4$\cite{Delaire_Na3PS4}, Cu$_7$PSe$_6$\cite{Delaire_Cu7PSe6} and AgCrSe$_2$\cite{Delaire_AgCrSe2}. To generate the training data, we have run first-principle molecular dynamics (AIMD) of the conventional unitcell for 120 ps at 1000K using the PWscf code of the Quantum ESPRESSO distribution \cite{Giannozzi_2017} and the Standard Solid-State Pseudopotential (SSSP) PBEsol efficiency 1.2 \cite{Prandini,Lejaeghere_SSSP} that verifies pseudopotentials from different methods and libraries \cite{VANSETTEN201839, SCHLIPF201536, TOPSAKAL2014263,DALCORSO2014337,NC_pseudo, GARRITY2014446}. Based on the results from defect formation energy calculations, we have selected Mg for substituion on Li-sublattice and Cl on S-sublattice. We have included 9 compositions in our training set (Li$_{7-x-y}$Mg$_x$PS$_{6-y}$Cl$_y$; x=0, 0.5, 1.5 and y=0, 0.25, 0.5) and we sampled a configuration every 6 fs. To investigate the sensitivity of the potential with respect to the size of the training set, we have also trained the potential using smaller datasets containing only 3 and 6 compositions. As can be seen in Fig. \ref{fig:training_set_convergence} of the Supplementary Information (SI), the ionic conductivity already `converges' after training with only 6 compositions. We also note that the potential trained with only 3 compositions shows some instabilities for some compositions that are not in the training, especially at 1250K which the highest temperature of our DeePMD simulations. The diffusivities of Li ions were computed from the slope of the mean-square displacements vs time and then converted into ionic conductivities using the Nernst-Einstein relation \cite{he_statistical_2018}  The initial configuration of the argyrodite was generated in such a way that the occupancies of Cl and S at zero Mg content on the octahedral and tetrahedral sites are as close as possible to the experimental values obtained from neutron diffraction \cite{gautam_Li7PS6} (see Fig. \ref{fig:Cl_occupancy}). Due to the lack of experimentally determined occupancies for Cl and S in the presence of other dopants, we assume that they do not change with the concentration of cation dopants.

\begin{figure*}
\centering
\includegraphics[width=1.0\linewidth]{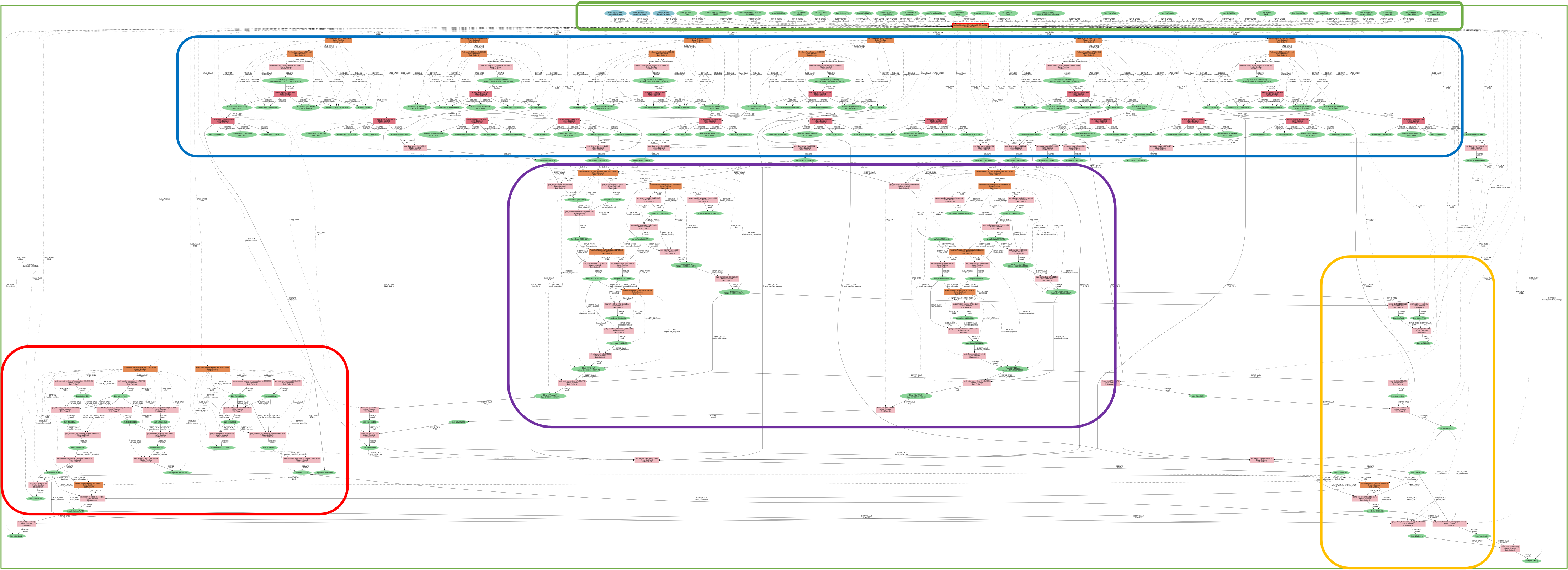}
% \vspace{1cm}
    \caption{An example of the provenance graph generated by AiiDA-defects for a typical series of defect calculations showing the inputs (green box) and outputs of the calculations along all the intermediate steps. All the nodes associated with the DFT calculations are in the blue box while the purple box contains all the nodes corresponding to the calculations of the correction terms. The red box contain nodes coming from the calculation of the chemical potentials and the orange box the self-consistent Fermi level.} 
    \label{fig:provenance}
\end{figure*}

As for the DeePMD training set, all DFT calculations for the defect formation energies in this work were carried out using the Quantum ESPRESSO distribution \cite{Giannozzi_2017} with PBESol \cite{PBESol_Perdew} as the exchange–correlation functional and the pseudopotentials and energy cutoffs taken from the SSSP PBEsol Efficiency 1.2 library \cite{Prandini,Lejaeghere_SSSP}. All calculations were performed on the 112 atom 2x2x2 supercell constructed from the primitive cell of the HT cubic phase of Li$_7$PS$_6$. The partial occupancy on the Li sites was removed using the Ewald criterion as implemented in the Pymatgen library \cite{ONG2013314}. All calculations for defect formation energies use a 2x2x2 Monkhorst-Pack kpoint mesh while for the AIMD, only the $\Gamma$ point was included.
To calculate the formation energies of various defects in the this study, we employ the AiiDA-defects package \cite{Muy_2023} which leverages the advanced capabilities of  AiiDA \cite{AiiDA}, an open-source materials informatics infrastructure that
provides workflow automation while simultaneously preserving and storing the full data provenance
in a relational database that is queryable, thus ensuring the reproducibility of the calculations. In AiiDA-defects, each term in Eq.\ref{eq:FormationEnergy} is computed by a dedicated AiiDA workchain which automatically generates inputs and gathers results of the calculations and stores them in a node which can be later readily queried for further analysis. An example of the provenance graph generated by AiiDA-defects for a small set of  defect formation energy calculations is shown in Fig. \ref{fig:provenance} and clearly illustrates how much automation is achieved.

\begin{figure}
% \centering
\includegraphics[width=0.95\linewidth]{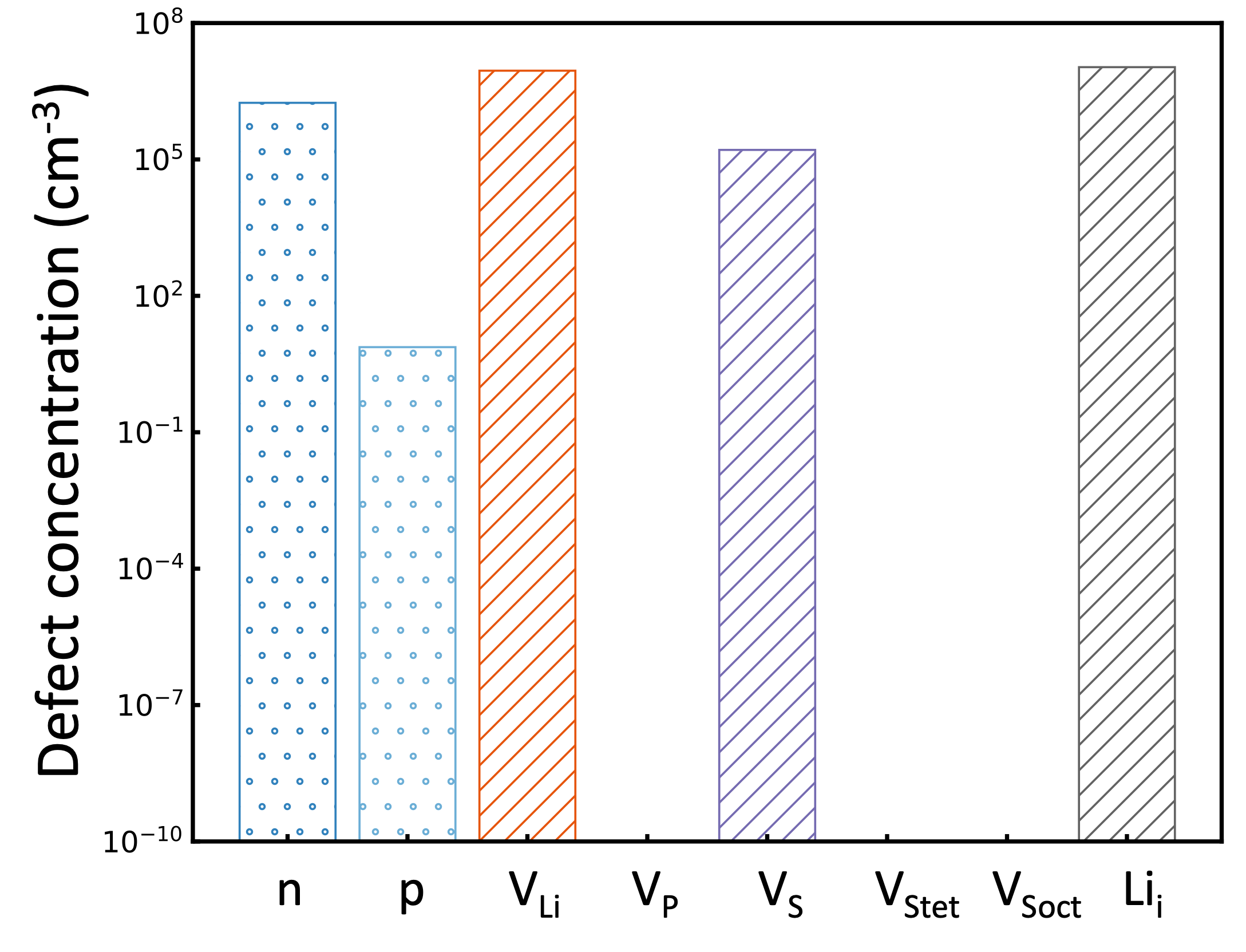}
    \caption{Concentrations of native defects in Li$_7$PS$_6$. The dominant defects in thermodynamic equilibrium are Li vacancies and interstitials. There is also a comparable concentration of electrons, suggesting that stoichiometric Li$_7$PS$_6$ exhibits a \emph{n}-type electronic conductivity in addition to Li-ion conductivity.} 
    \label{fig:intrinsic_defect}
\end{figure}

\begin{figure*}[t]
\centering
\includegraphics[width=0.95\linewidth]{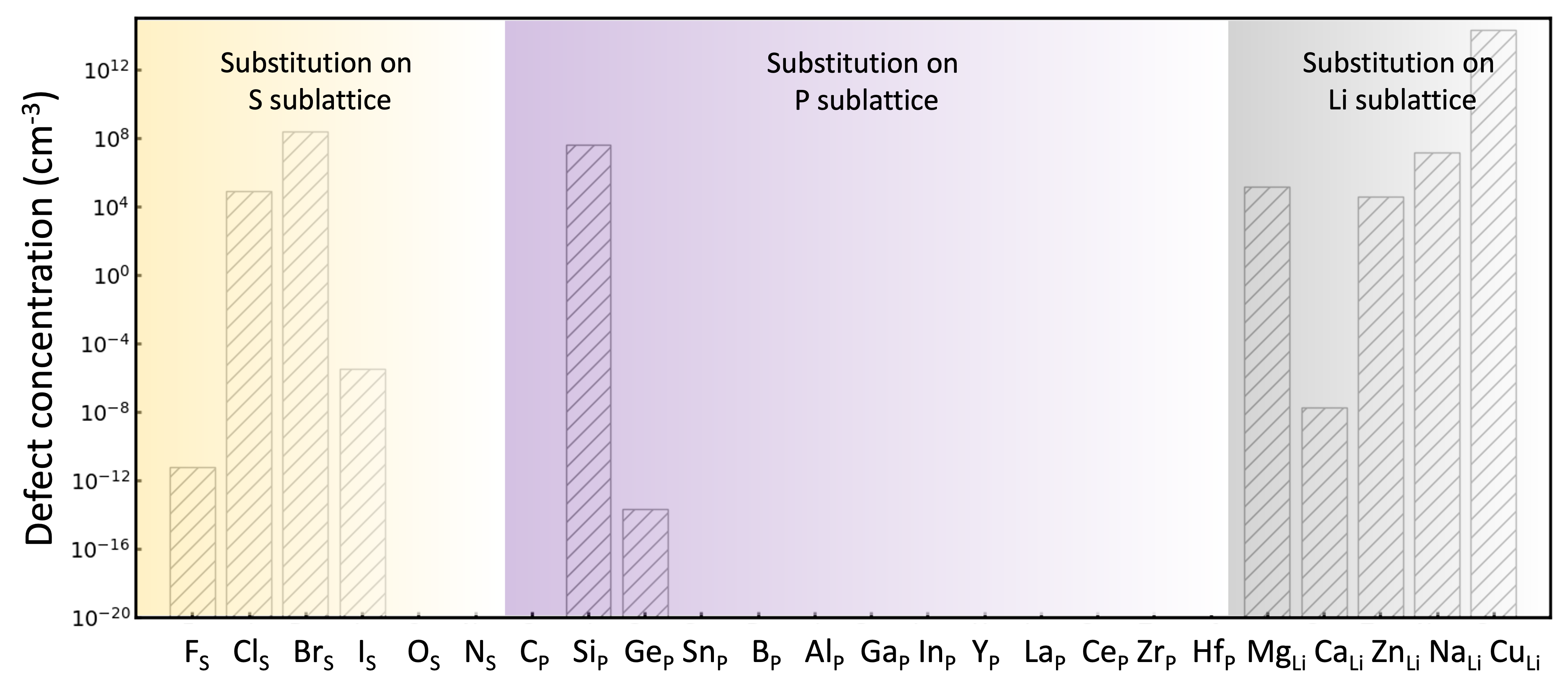}
% \vspace{1cm}
    \caption{Concentrations of isovalent and aliovalent substitutions on the Li, P and S sub-lattices in the Li$_7$PS$_6$ argyrodite.} 
    \label{fig:extrinsic}
\end{figure*}

\begin{figure*}
\centering
\includegraphics[width=1.0\linewidth]{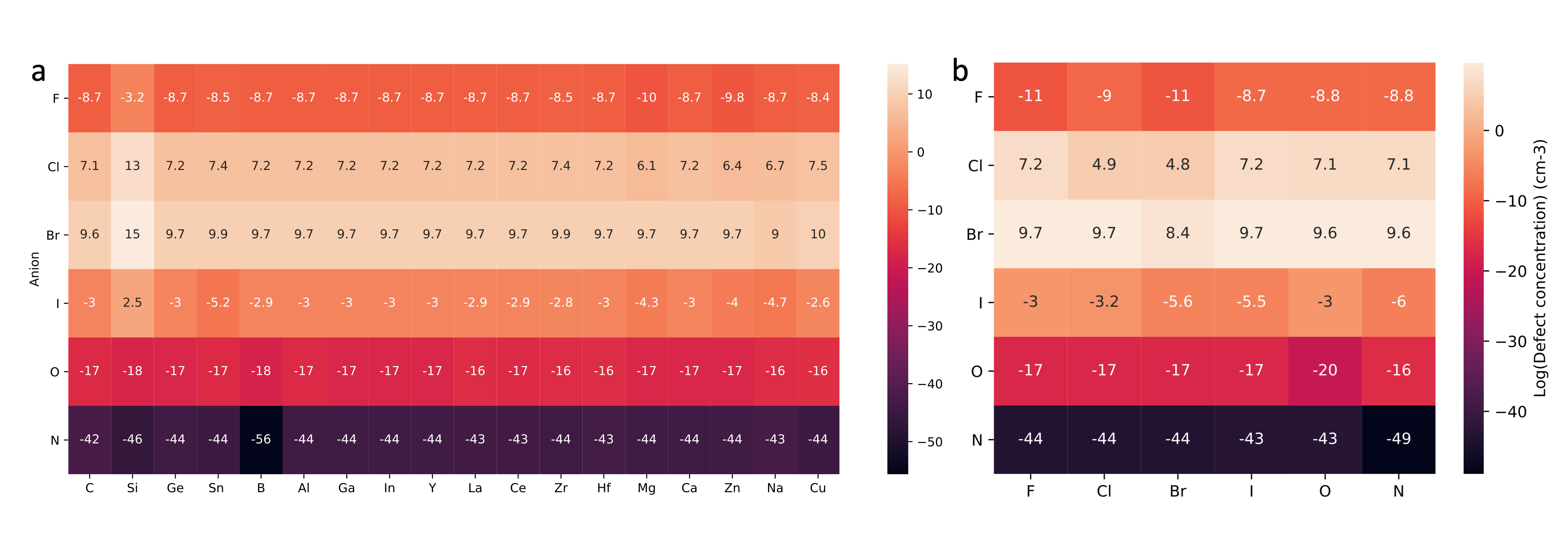}
% \vspace{1cm}
    \caption{a) Concentration of anion substitutions on S sites as function of cation co-dopants. Generally speaking, the presence of cations as dopants doesn't affect the concentration of anion dopants in any significant way, except the case of Si co-dopant where the concentration of the anion can be increased substantially. b) Concentration of anion substitutions on S sites as a function of other anion co-dopants. In general, it can be seen that the presence of other anion co-dopants, especially within the halogen family, can significantly alter the concentration of each anion in Li$_7$PS$_6$ argyrodites.} 
    \label{fig:codopant}
\end{figure*}

\section{\label{sec:section3}Results and discussions}
\subsection{\label{sec:section3.1}Native defects}
The concentrations of native defect in the HT phase of Li$_7$PS$_6$ at 300K are shown in Figure \ref{fig:intrinsic_defect}. The dominant ionic defects are predicted to be lithium vacancies and interstitials. There is also comparable concentration of electrons, suggesting that stoichiometric Li$_7$PS$_6$ exhibits a \emph{n}-type electronic conductivity in addition to Li-ion conductivity. Notice that this is the concentration of defects corresponding to the `typical' chemical potential of each element as given by the centroid of the stability region \cite{BUCKERIDGECHEMICALPOTENTIAL}. These concentrations are expected to change when the chemical potentials are chosen from a different point in the stability region that corresponds for instance to more S-rich or S-poor conditions.

\subsection{\label{sec:section3.2}Isovalent and aliovalent substitutions}
In the following section, we discuss the doping response of Li$_7$PS$_6$ to isovalent and aliovalent substitutions on Li, P and S sites.
\subsubsection{\label{sec:section3.2.1}Anion substitutions on S sites}
\begin{figure*}[t]
\centering
\includegraphics[width=0.95\linewidth]{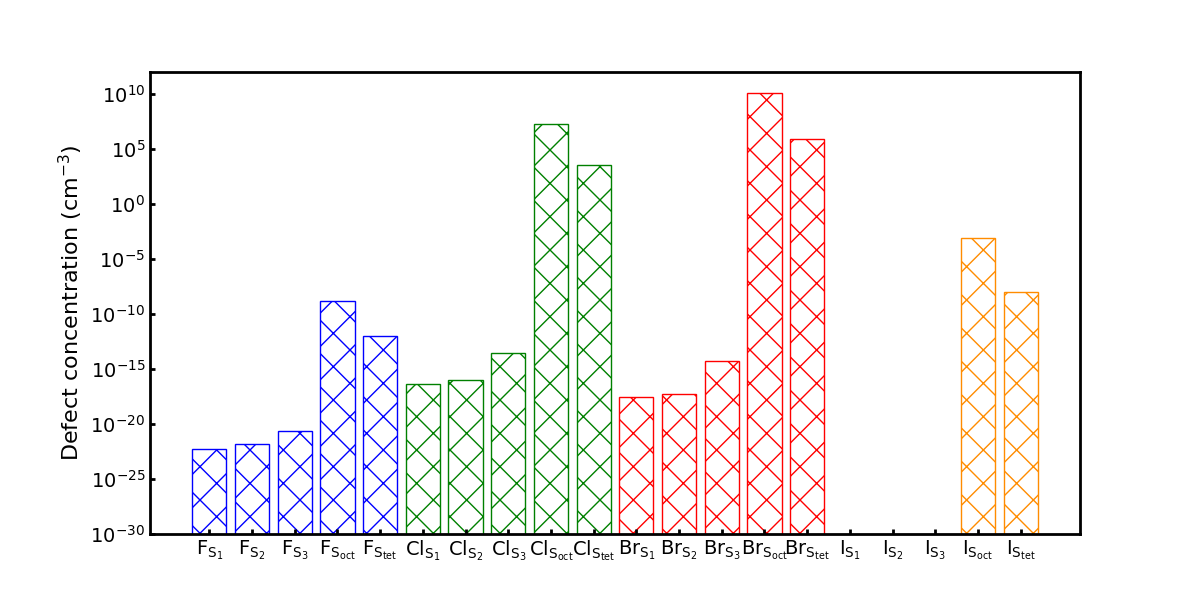}
% \vspace{1cm}
    \caption{Concentration of halide substitutions on various sulfur sites in cubic argyrodite Li$_7$PS$_6$. The concentration of F at equilibrium is very low compared to other halide substitutions, consistent with the absence of reports of F-doped Li argyrodites. On the other hand, Cl, Br and I readily substitute S on the octahedral and tetahedral sites, in agreement with experimental findings.} 
    \label{fig:halide_dopant}
\end{figure*}

The computed concentration of halide substitutions on the S sites are shown in Fig. \ref{fig:extrinsic} (more detailed breakdown of the concentrations of halide substitutions on the each non-equivalent S sites are shown in Figure \ref{fig:halide_dopant}). It can be seen clearly from Figure \ref{fig:extrinsic} that Cl, Br and I substitutions are much more favorable than F substitution in agreement with the apparent absence of F-substituted Li-argyrodites in the litterature \cite{Ohno_review, muy_review, YU2021105858}. Moreover, the most favorable sulfur sites for halide substitutions are the octahedral (4a) and tetrahedral (4d) sites, as opposed to the 16d sites in which the S ions are bonded to the phosphorus ions (Fig. \ref{fig:halide_dopant}), also in agreement with experiments. Finally, we note that our calculations suggest that at room-temperature the concentrations of halide substitutions on the octahedral sites are almost 5 orders of magnitude larger than that on the tetrahedral sites which implies that at RT, the halides (Cl$^{-}$, Br$^{-}$ and I$^{-}$) anions will only occupy the octahedral site. Experimentally, this has been shown to be indeed the case for I substitutions but not for Cl and Br substitutions where it is well established that they can both occupy tet and oct sites along with S ions leading to a disorder on the sublattice which can greatly enhance the long-range diffusion of Li-ions \cite{deKlerk, kraft_influence_2017, Kraft, Morgan, Stamminger, RayavarapuMAY}. This apparent discrepancy might point to the fact that the actual partial occupancy of Cl and Br on oct/tet site at RT is actually a metastable structure which is kinetically trapped from the higher temperatures at which these compounds are synthesized. This hypothesis is consistent with the fact that thermal treatments have been used to optimize the occupancy of S and Cl/Br on the oct/tet sites to achieve higher ionic conductivity \cite{Gautam}.

\subsubsection{\label{sec:section3.2.2}Cation substitutions on Li and P sites}
To study the effect of doping response to cation substitutions, we have considered some of the most chemically plausible mono-, di-, tri- and tetra-valent cations on the Li and P sites. The concentrations of these substitutions as determined from the defect formation energies are displayed in Figure. \ref{fig:extrinsic}. 
For the isovalent substitutions, the calculations show that Na$_{\text{Li}}$ and Cu$_{\text{Li}}$ are energetically very favorable, which is consistent with the presence of Ag argyrodite analogue \cite{Eulenberger1977, Hahn1965}. Divalent dopants such as Mg$^{2+}$ and Zn$^{2+}$ can also readily replace Li$^+$ ions, suggesting another less explored strategy to fine-tune the concentration of Li and therefore the its ionic conductivity. A major downside of this approach is that at high concentrations these ions might co-diffuse with Li, which can be detrimental for battery operation. The critical concentrations of dopants at which this co-diffusion becomes non-negligible requires further investigations. Ca$^{2+}$ is predicted to be less favorable than Mg$^{2+}$ and Zn$^{2+}$ but could still be doped at practically relevant concentration. This is indeed in agreement with the reported synthesis of Li$_{5.35}$Ca$_{0.1}$PS$_{4.5}$Cl$_{1.55}$ with ionic conductivities of 10.2 mS/cm at RT \cite{Adeli_Ca_doped}. 
%Another particularity of Ca$^{2+}$ is that unlike Mg$^{2+}$ and Zn$^{2+}$ which can replace Li at all sites, Ca$^{2+}$ seems to prefer a particular site of Li, suggesting a possible ordering of Ca sublattice.
Al$^{3+}$ substitution on Li sites is predicted to be very unfavorable, which seems to be in contradiction with the report synthesis of Li$_{6.15}$Al$_{0.15}$Si$_{1.55}$S$_6$ \cite{HUANG2019487}, suggesting again that this composition corresponds to a metastable structure rather than a thermodynamically stable one. Finally, among all the tri- and tetravalent substitutions on P-sites considered in this study, only Si and Ge show favorable energetics at thermodynamic equilibrium, consistent with the reports of successful synthesis of Li$_{7+x}$P$_{1-x}$M$_x$S$_6$ with M = Si and Ge; $x \approx 0.3-0.35$. \cite{SCHNEIDER2017151,C8TA10790D}

\subsubsection{\label{sec:section3.2.3}Cation-anion co-dopings}
% \begin{figure*}[t]
% \centering
% \includegraphics[width=1.0\linewidth]{images/Divalent_dopant_300K.png}
% % \vspace{1cm}
%     \caption{Concentration of divalent substitutions Mg$^{2+}$, Ca$^{2+}$ and Zn$^{2+}$ on Li sites in thermodynamic equilibrium. Mg and Zn readily substitute Li on every non-equivalent crystallographic sites. On the other hand, Ca shows a moderate tendency to replace Li with a strong preference for a particular Li sites, suggesting a possible ordering of Ca in Li$_7$PS$_6$ argyrodite.} 
%     \label{fig:divalent_dopant}
% \end{figure*}

Several compounds in the argyrodite family that show high ionic conductivity at RT are actually obtained from the parent compound Li$_7$PS$_6$ by co-substitution of both cation and anion as in,  for example, Li$_{6+x}$P$_{1-x}$M$_x$S$_5$I (M=Si, Ge) \cite{Kraft}. It is therefore of great interest to study these co-doping scenarios in detail in order to find novel pairs of co-dopants which are thermodynamically favorable. The formalism used in previous sections to study a single dopant can be readily extended to include two or even more dopants. The main assumption here is that the defects do not interact directly with one another and that the presence of other defects is mediated only through the charge-neutrality condition imposed in the calculation of the Fermi level. With this assumption, one doesn't have to compute the energy of the supercell with two dopants simultaneously, only the (DFT) energy of supercells with single dopants is required and the effect of co-dopants is entirely captured in the change of the self-consistent Fermi level as determined by Eq. \ref{eq:ChargeNeutrality}. The concentrations of halide, oxygen and nitrogen ions as a function of cation co-dopants are shown in Figure \ref{fig:codopant}a. One can see that in general the presence of cation co-dopant doesn't affect the concentration of anion dopants in any significant way, except the case of Si co-dopant where the concentration of the anion can be increased substantially, raising the prospect that one might be able to synthesize the first F-doped argyrodite by co-doping the compound with Si. The results for the anion-anion co-doping are shown in Figure \ref{fig:codopant}b. Unlike cation co-doping, the presence of other anion co-dopants, especially within the halogen family, can significantly alter the concentration of the other anion in Li$_7$PS$_6$ argyrodites.

\begin{figure*}
\centering
\includegraphics[width=1.\linewidth]{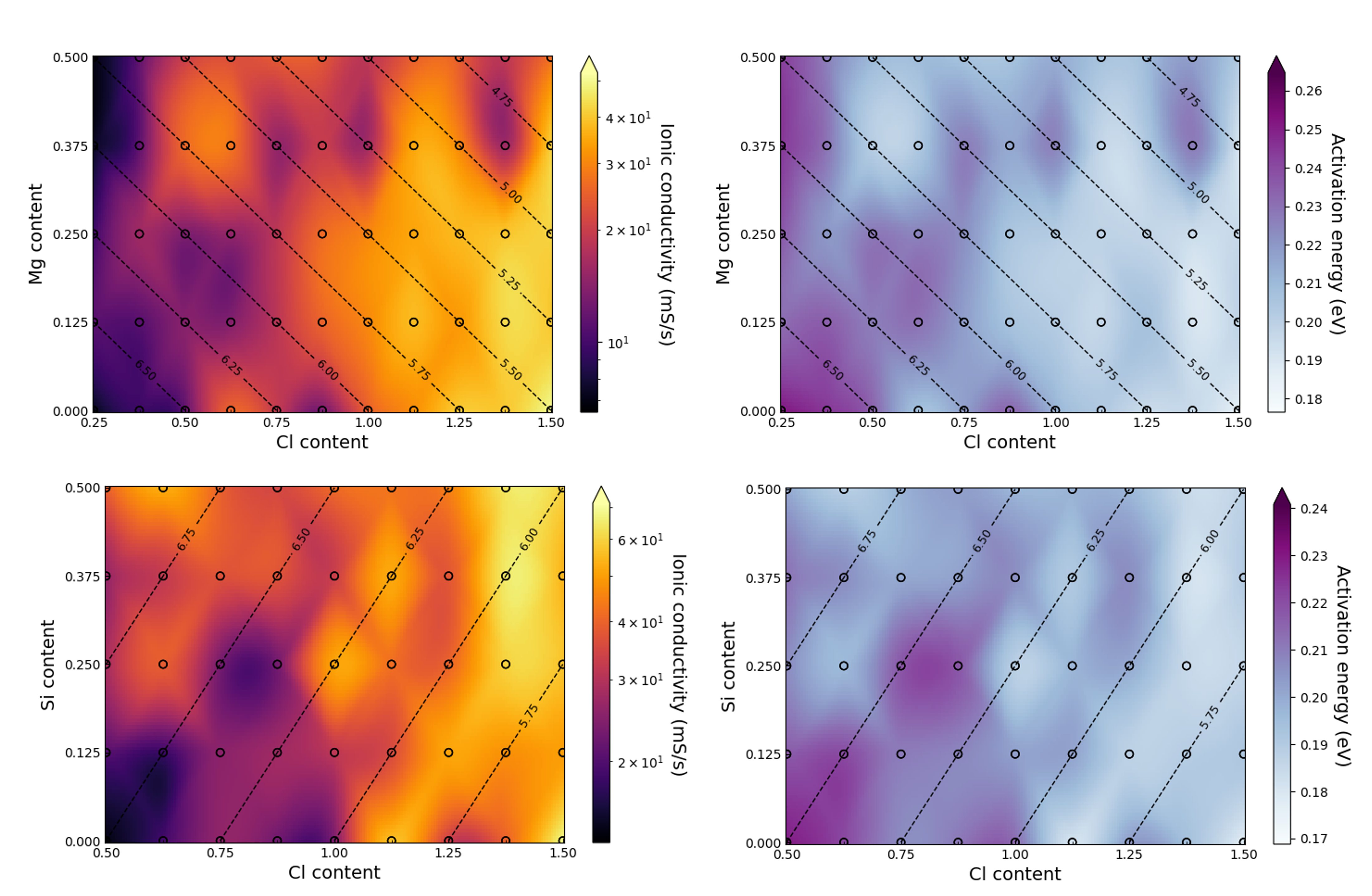}
% \vspace{1cm}
    \caption{The ionic conductivity and activation energy landscape of Mg-Cl and Si-Cl co-doped Li$_7$PS$_6$ argyrodite as computed using the DeePMD potential \cite{DeePMD_kit, DeePMD_SE}. The dashed lines corresponds to the compositions with the same Li content. The region of the compositional space with the highest ionic conductivity (lowest activation energy) are the ones with high Cl content and low Mg/Si contents.} 
    \label{fig:sigma_Ea_landscape}
\end{figure*}

\subsection{\label{sec:section3.3}Ionic conductivity landscape}
\begin{figure*}[t]
\centering
\includegraphics[]{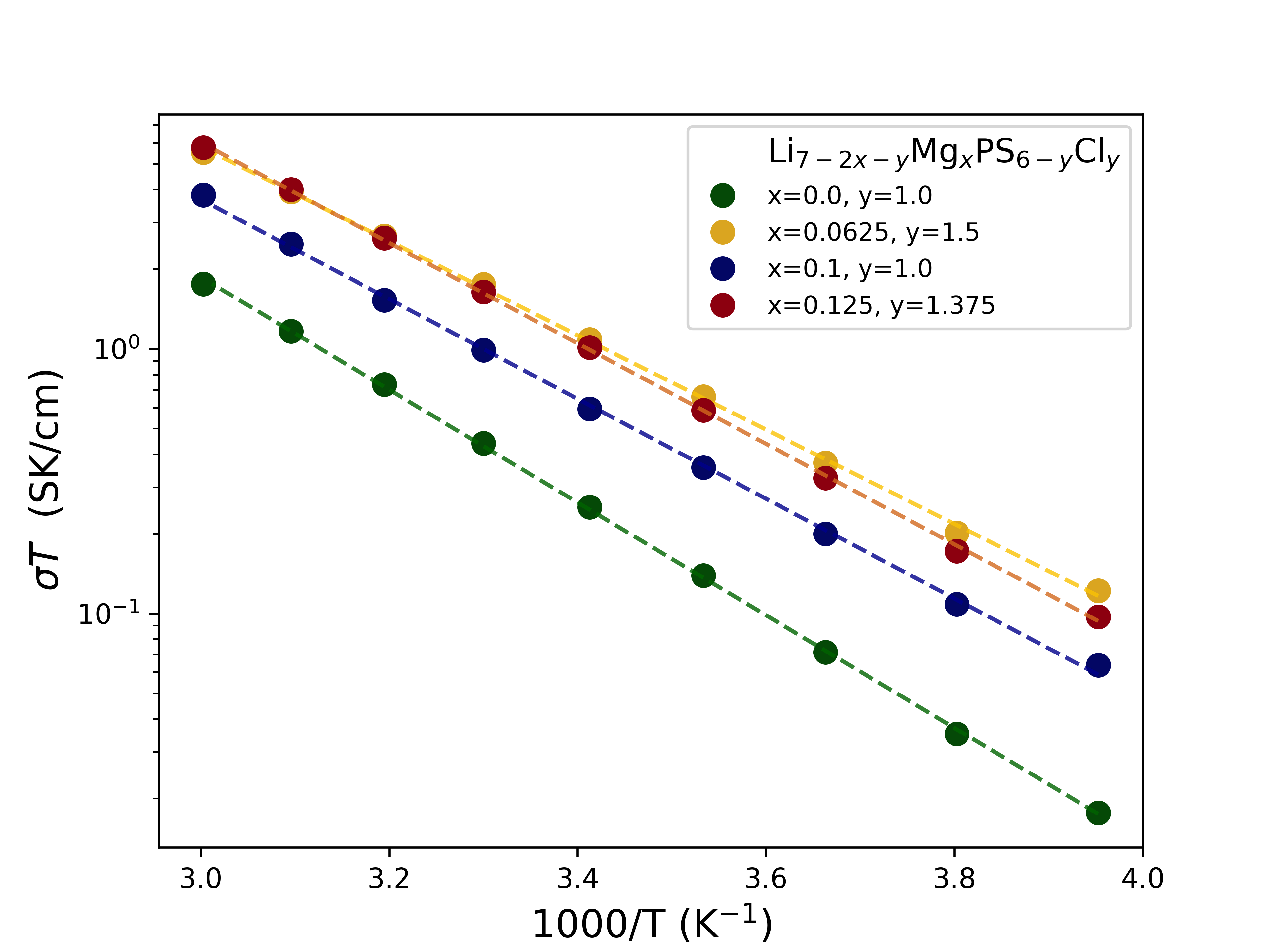}
% \vspace{1cm}
    \caption{Experimental ionic conductivity of Li in Mg-doped Argyrodite Li$_7-2x-y$Mg$_x$PS$_6-y$Cl$_y$ measured between -20 and 60$^{o}$C, showing good agreement with the computed results.} 
    \label{fig:arrhenius_plot}
\end{figure*}

\begin{figure}
    \centering
    \subfigure[]{\includegraphics[width=0.475\textwidth]{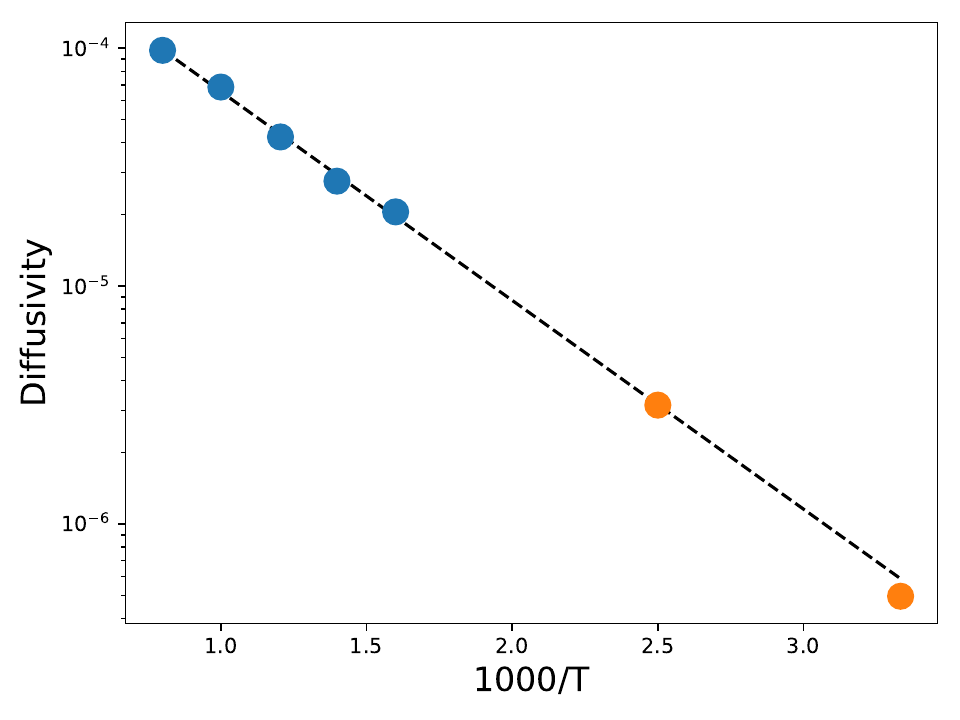}} 
    \subfigure[]{\includegraphics[width=0.475\textwidth]{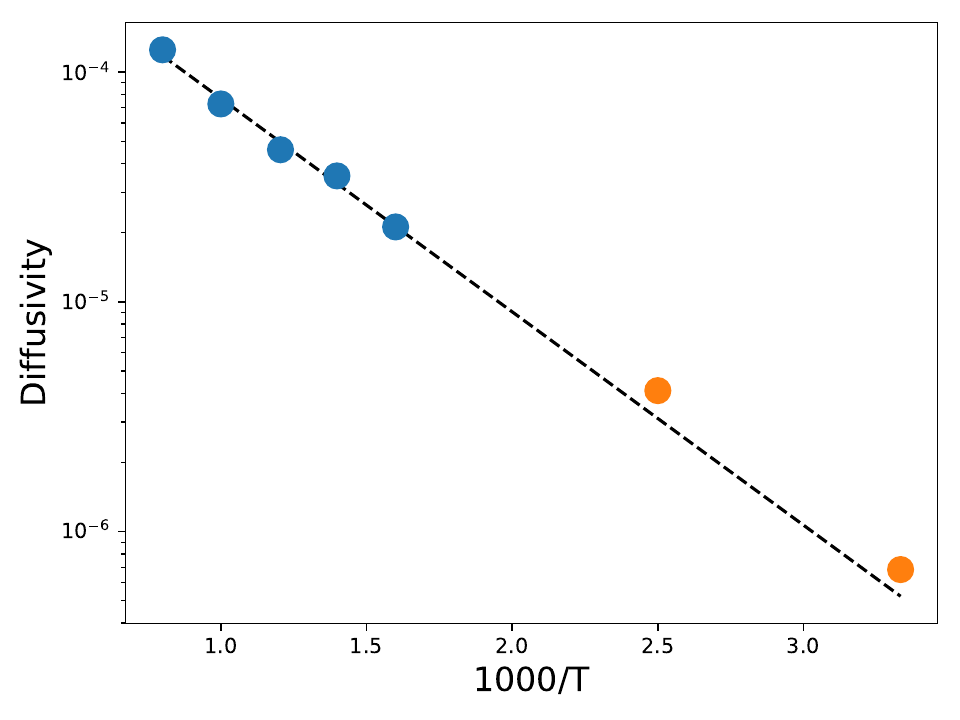}} 
    \caption{Comparison between the computed and extrapolated diffusivity of Li in a) Li$_{5.75}$Mg$_{0.375}$PS$_{4.5}$Cl$_{0.5}$ and b) Li$_{5.375}$Mg$_{0.25}$PS$_{4.875}$Cl$_{1.125}$.}
    \label{fig:extrapolated_sigma}
\end{figure}

Based on the calculations of defect formation energies, Mg, Si and Cl are predicted to be some of the most promising dopants on Li, P and S sub-lattices respectively. In order to investigate the effects of these dopants on the ionic conductivity of Li-ion in the (cubic) argyrodite structure, we have trained several DeePMD potentials \cite{DeePMD_kit, DeePMD_SE} using the protocol previously described in Section \ref{sec:section2} to run classical MD of the argyrodite over a wide range of dopant concentrations. The ionic conductivities at 300K and the activation energies for Mg-Cl and Si-Cl co-doping are shown in Fig. \ref{fig:sigma_Ea_landscape}. The ionic conductivities are extrapolated from higher temperatures (1250 K, 1000 K, 840 K, 715 K and 625 K) to room temperature, using Arrhenius equation. We have verified for selected compositions that the extrapolated and the computed values are in very good agreement (see Fig. \ref{fig:extrapolated_sigma}). The lower bound of Cl is set to 0.25 and 0.5 respectively for Mg and Si co-doping scenario, as it is known that the cubic phase of pure Li$_7$PS$_6$ is not stable at room temperature. The upper bound of Cl is set to 1.5 as it is was found that  secondary phases start to form beyond this limit \cite{gautam_Li7PS6}. While there are several local maxima/minima at some particular compositions, in both cases it is clear that the regions of the compositional space with the highest ionic conductivity (lowest activation energy) are those with high Cl content and low Mg/Si content. 
Based on this information, we have synthesized several (Mg,Cl)-doped argyrodites Li$_{6-2x}$Mg$_x$PS$_{6-y}$Cl$_y$ with x=0, 0.125, 0.0625, 0.1 and 0.25; y = 1, 1.375 and 1.5. All compositions were found to be phase-pure except for Li$_{5.25}$Mg$_{0.25}$PS$_{4.75}$Cl$_{1.25}$ where trace of MgS was detected. The measured ionic conductivity of these samples between -20 and 60$^{o}$C are shown in Fig. \ref{fig:arrhenius_plot} and were in good agreement with the computed results.
The activation energies of Li conduction were found to be 0.38 eV for Li$_6$PS$_5$Cl, 0.38 eV for Li$_{5.375}$Mg$_{0.125}$PS$_{4.625}$Cl$_{1.375}$, 0.36 eV for Li$_{5.375}$Mg$_{0.0625}$PS$_{4.5}$Cl$_{1.5}$ and 0.38 eV for Li$_{5.8}$Mg$_{0.1}$PS$_5$Cl. RT Electronic conductivity for all samples is smaller than $10^{-8}$ S/cm \cite{patent_1,patent_2}. These experimental measurements demonstrate the effectiveness of computational materials design as a complementary tool to accelerate the discovery of new functional materials with improved properties.

\section{Conclusion}
In this study, we have provided a thorough description of the intrinsic defects as well as dopant regimes in the Li argyrodite Li$_7$PS$_6$ using the AiiDA-defects package to enable the calculation of defect formation energies in a robust, reproducible and thermodynamically consistent way. We found that halide dopants, with the exception of iodine readily substitute on the octahedral and tetrahedral sites, in agreement with the experiments. We also find that Mg, Zn, Si and Ge are some of the most promising dopants to vary the concentration of Li-ion and its ionic conductivity, while tri-valent dopants are predicted to be thermodynamically unfavorable. We have also trained DeePMD potentials to perform MD simulations to investigate the effects of these dopants on the ionic conductivities of Li, by establishing an ionic conductivity map that indicates that the highest Li-ion conductivity can be achieved by having high concentrations of Cl but low concentration of Mg/Si dopants. This approach provides valuable information to guide the synthesis and optimization of new argyrodite compositions with higher ionic conductivity.

\section*{Acknowledgement}
The authors would like to thank the Swiss National Supercomputing Centre CSCS (project s1073) for providing the computational ressources and Syensqo for their financial and scientific supports for this project.

% \clearpage
\bibliography{biblio}

\begin{thebibliography}{10}

\bibitem{Adeli_Ca_doped}
{\sc Adeli, P., Bazak, J.~D., Huq, A., Goward, G.~R., and Nazar, L.~F.}
\newblock Influence of aliovalent cation substitution and mechanical compression on li-ion conductivity and diffusivity in argyrodite solid electrolytes.
\newblock {\em Chemistry of Materials 33}, 1 (2021), 146--157.

\bibitem{muy_review}
{\sc Bachman, J.~C., Muy, S., Grimaud, A., Chang, H.-H., Pour, N., Lux, S.~F., Paschos, O., Maglia, F., Lupart, S., Lamp, P., Giordano, L., and Shao-Horn, Y.}
\newblock Inorganic solid-state electrolytes for lithium batteries: Mechanisms and properties governing ion conduction.
\newblock {\em Chemical Reviews 116}, 1 (2016), 140--162.

\bibitem{Bernges}
{\sc Bernges, T., Culver, S.~P., Minafra, N., Koerver, R., and Zeier, W.~G.}
\newblock Competing structural influences in the li superionic conducting argyrodites li6ps5–xsexbr (0 $\leq$ x $\leq$ 1) upon se substitution.
\newblock {\em Inorganic Chemistry 57}, 21 (2018), 13920--13928.

\bibitem{BUCKERIDGEFERMI}
{\sc Buckeridge, J.}
\newblock Equilibrium point defect and charge carrier concentrations in a material determined through calculation of the self-consistent fermi energy.
\newblock {\em Computer Physics Communications 244\/} (2019), 329 -- 342.

\bibitem{BUCKERIDGECHEMICALPOTENTIAL}
{\sc Buckeridge, J., Scanlon, D., Walsh, A., and Catlow, C.}
\newblock Automated procedure to determine the thermodynamic stability of a material and the range of chemical potentials necessary for its formation relative to competing phases and compounds.
\newblock {\em Computer Physics Communications 185}, 1 (2014), 330 -- 338.

\bibitem{ChenC6MH00218H}
{\sc Chen, R., Qu, W., Guo, X., Li, L., and Wu, F.}
\newblock The pursuit of solid-state electrolytes for lithium batteries: from comprehensive insight to emerging horizons.
\newblock {\em Mater. Horiz. 3\/} (2016), 487--516.

\bibitem{dabo_electrostatics_2008}
{\sc Dabo, I., Kozinsky, B., Singh-Miller, N.~E., and Marzari, N.}
\newblock Electrostatics in periodic boundary conditions and real-space corrections.
\newblock {\em Physical Review B 77}, 11 (Mar. 2008), 115139.

\bibitem{DALCORSO2014337}
{\sc {Dal Corso}, A.}
\newblock Pseudopotentials periodic table: From h to pu.
\newblock {\em Computational Materials Science 95\/} (2014), 337--350.

\bibitem{deKlerk}
{\sc de~Klerk, N. J.~J., Rosłoń, I., and Wagemaker, M.}
\newblock Diffusion mechanism of li argyrodite solid electrolytes for li-ion batteries and prediction of optimized halogen doping: The effect of li vacancies, halogens, and halogen disorder.
\newblock {\em Chemistry of Materials 28}, 21 (2016), 7955--7963.

\bibitem{Deiseroth}
{\sc Deiseroth, H.-J., Kong, S.-T., Eckert, H., Vannahme, J., Reiner, C., Zai{\ss}, T., and Schlosser, M.}
\newblock Li6ps5x: A class of crystalline li-rich solids with an unusually high li+ mobility.
\newblock {\em Angewandte Chemie International Edition 47}, 4 (2008), 755--758.

\bibitem{Deiseroth_Li7PS6}
{\sc Deiseroth, H.-J., Maier, J., Weichert, K., Nickel, V., Kong, S.-T., and Reiner, C.}
\newblock Li7ps6 and li6ps5x (x: Cl, br, i): Possible three-dimensional diffusion pathways for lithium ions and temperature dependence of the ionic conductivity by impedance measurements.
\newblock {\em Zeitschrift für anorganische und allgemeine Chemie 637}, 10 (2011), 1287--1294.

\bibitem{Delaire_AgCrSe2}
{\sc Ding, J., Niedziela, J.~L., Bansal, D., Wang, J., He, X., May, A.~F., Ehlers, G., Abernathy, D.~L., Said, A., Alatas, A., Ren, Y., Arya, G., and Delaire, O.}
\newblock Anharmonic lattice dynamics and superionic transition in agcrse$_2$.
\newblock {\em Proceedings of the National Academy of Sciences 117}, 8 (2020), 3930--3937.

\bibitem{patent_1}
{\sc EMERY, A., BRAIDA, M.-D., Mercier, T.~L., Marzari, N., MUY, S., and ERCOLE, L.}
\newblock Solid material comprising li, mg, p, s and halogen elements, jun 2023.

\bibitem{patent_2}
{\sc EMERY, A., BRAIDA, M.-D., Mercier, T.~L., Marzari, N., MUY, S., and ERCOLE, L.}
\newblock Solid material comprising li, mg, p, s and halogen elements, jun 2023.

\bibitem{Eulenberger1977}
{\sc Eulenberger, G.}
\newblock Die kristallstruktur der tieftemperaturmodifikation von ag8ges6.
\newblock {\em Monatshefte f{\"u}r Chemie / Chemical Monthly 108}, 4 (Jul 1977), 901--913.

\bibitem{FreysoldtRMP}
{\sc Freysoldt, C., Grabowski, B., Hickel, T., Neugebauer, J., Kresse, G., Janotti, A., and Van~de Walle, C.~G.}
\newblock First-principles calculations for point defects in solids.
\newblock {\em Rev. Mod. Phys. 86\/} (Mar 2014), 253--305.

\bibitem{freysoldt_PRL_2009}
{\sc Freysoldt, C., Neugebauer, J., and Van~de Walle, C.~G.}
\newblock Fully {Ab} {Initio} {Finite}-{Size} {Corrections} for {Charged}-{Defect} {Supercell} {Calculations}.
\newblock {\em Physical Review Letters 102}, 1 (Jan. 2009), 016402.

\bibitem{freysoldt_2011}
{\sc Freysoldt, C., Neugebauer, J., and Walle, C. G. V.~d.}
\newblock Electrostatic interactions between charged defects in supercells.
\newblock {\em physica status solidi (b) 248}, 5 (2011), 1067--1076.

\bibitem{GARRITY2014446}
{\sc Garrity, K.~F., Bennett, J.~W., Rabe, K.~M., and Vanderbilt, D.}
\newblock Pseudopotentials for high-throughput dft calculations.
\newblock {\em Computational Materials Science 81\/} (2014), 446--452.

\bibitem{gautam_Li7PS6}
{\sc Gautam, A., Ghidiu, M., Suard, E., Kraft, M.~A., and Zeier, W.~G.}
\newblock On the lithium distribution in halide superionic argyrodites by halide incorporation in li$_{7–x}$ps$_{6–x}$cl$_x$.
\newblock {\em {ACS} Applied Energy Materials 4}, 7 (2021), 7309--7315.
\newblock Publisher: American Chemical Society.

\bibitem{Gautam}
{\sc Gautam, A., Sadowski, M., Prinz, N., Eickhoff, H., Minafra, N., Ghidiu, M., Culver, S.~P., Albe, K., Fässler, T.~F., Zobel, M., and Zeier, W.~G.}
\newblock Rapid crystallization and kinetic freezing of site-disorder in the lithium superionic argyrodite li6ps5br.
\newblock {\em Chemistry of Materials 31}, 24 (2019), 10178--10185.

\bibitem{Giannozzi_2017}
{\sc Giannozzi, P., Andreussi, O., Brumme, T., Bunau, O., Nardelli, M.~B., Calandra, M., Car, R., Cavazzoni, C., Ceresoli, D., Cococcioni, M., Colonna, N., Carnimeo, I., Corso, A.~D., de~Gironcoli, S., Delugas, P., DiStasio, R.~A., Ferretti, A., Floris, A., Fratesi, G., Fugallo, G., Gebauer, R., Gerstmann, U., Giustino, F., Gorni, T., Jia, J., Kawamura, M., Ko, H.-Y., Kokalj, A., K{\"{u}}{\c{c}}{\"{u}}kbenli, E., Lazzeri, M., Marsili, M., Marzari, N., Mauri, F., Nguyen, N.~L., Nguyen, H.-V., de-la Roza, A.~O., Paulatto, L., Ponc{\'{e}}, S., Rocca, D., Sabatini, R., Santra, B., Schlipf, M., Seitsonen, A.~P., Smogunov, A., Timrov, I., Thonhauser, T., Umari, P., Vast, N., Wu, X., and Baroni, S.}
\newblock Advanced capabilities for materials modelling with quantum {ESPRESSO}.
\newblock {\em Journal of Physics: Condensed Matter 29}, 46 (oct 2017), 465901.

\bibitem{Delaire_Cu7PSe6}
{\sc Gupta, M.~K., Ding, J., Bansal, D., Abernathy, D.~L., Ehlers, G., Osti, N.~C., Zeier, W.~G., and Delaire, O.}
\newblock Strongly anharmonic phonons and their role in superionic diffusion and ultralow thermal conductivity of cu7pse6.
\newblock {\em Advanced Energy Materials 12}, 23 (2022), 2200596.

\bibitem{Delaire_Na3PS4}
{\sc Gupta, M.~K., Ding, J., Osti, N.~C., Abernathy, D.~L., Arnold, W., Wang, H., Hood, Z., and Delaire, O.}
\newblock Fast na diffusion and anharmonic phonon dynamics in superionic na3ps4.
\newblock {\em Energy Environ. Sci. 14\/} (2021), 6554--6563.

\bibitem{Hahn1965}
{\sc Hahn, H., Schulze, H., and Sechser, L.}
\newblock {\"U}ber einige tern{\"a}re chalkogenide vom argyrodit-typ.
\newblock {\em Naturwissenschaften 52}, 15 (Jan 1965), 451--451.

\bibitem{he_statistical_2018}
{\sc He, X., Zhu, Y., Epstein, A., and Mo, Y.}
\newblock Statistical variances of diffusional properties from ab initio molecular dynamics simulations.
\newblock {\em npj Computational Materials 4}, 1 (2018), 1--9.

\bibitem{DeePMD_LGPS}
{\sc Huang, J., Zhang, L., Wang, H., Zhao, J., Cheng, J., and E, W.}
\newblock Deep potential generation scheme and simulation protocol for the li10gep2s12-type superionic conductors.
\newblock {\em The Journal of Chemical Physics 154}, 9 (2021), 094703.

\bibitem{HUANG2019487}
{\sc Huang, W., Yoshino, K., Hori, S., Suzuki, K., Yonemura, M., Hirayama, M., and Kanno, R.}
\newblock Superionic lithium conductor with a cubic argyrodite-type structure in the li–al–si–s system.
\newblock {\em Journal of Solid State Chemistry 270\/} (2019), 487--492.

\bibitem{AiiDA}
{\sc Huber, S.~P., Zoupanos, S., Uhrin, M., Talirz, L., Kahle, L., H{\"{a}}uselmann, R., Gresch, D., M{\"{u}}ller, T., Yakutovich, A.~V., Andersen, C.~W., Ramirez, F.~F., Adorf, C.~S., Gargiulo, F., Kumbhar, S., Passaro, E., Johnston, C., Merkys, A., Cepellotti, A., Mounet, N., Marzari, N., Kozinsky, B., and Pizzi, G.}
\newblock {AiiDA} 1.0, a scalable computational infrastructure for automated reproducible workflows and data provenance.
\newblock {\em Scientific Data 7}, 1 (Sept. 2020), 300.

\bibitem{Janek}
{\sc Janek, J., and Zeier, W.}
\newblock A solid future for battery development.
\newblock {\em Nature Energy 1}, 9 (2016), 16141 -- 16144.

\bibitem{Kong_Li6PO5Cl}
{\sc Kong, S.-T., Deiseroth, H.-J., Maier, J., Nickel, V., Weichert, K., and Reiner, C.}
\newblock Li6po5br and li6po5cl: The first lithium-oxide-argyrodites .
\newblock {\em Zeitschrift für anorganische und allgemeine Chemie 636}, 11 (2010), 1920--1924.

\bibitem{Kong_As}
{\sc Kong, S.-T., Deiseroth, H.-J., Reiner, C., G\"{u}n, O., Neumann, E., Ritter, C., and Zahn, D.}
\newblock Lithium argyrodites with phosphorus and arsenic: Order and disorder of lithium atoms, crystal chemistry, and phase transitions.
\newblock {\em Chemistry – A European Journal 16}, 7 (2010), 2198--2206.

\bibitem{Kong_Li7PS6}
{\sc Kong, S.-T., G\"{u}n, O., Koch, B., Deiseroth, H.-J., Eckert, H., and Reiner, C.}
\newblock Structural characterisation of the li argyrodites li7ps6 and li7pse6 and their solid solutions: Quantification of site preferences by mas-nmr spectroscopy.
\newblock {\em Chemistry – A European Journal 16}, 17 (2010), 5138--5147.

\bibitem{kraft_influence_2017}
{\sc Kraft, M.~A., Culver, S.~P., Calderon, M., Böcher, F., Krauskopf, T., Senyshyn, A., Dietrich, C., Zevalkink, A., Janek, J., and Zeier, W.~G.}
\newblock Influence of {Lattice} {Polarizability} on the {Ionic} {Conductivity} in the {Lithium} {Superionic} {Argyrodites} {Li}$_{\textrm{6}}${PS}$_{\textrm{5}}${X} ({X} = {Cl}, {Br}, {I}).
\newblock {\em Journal of the American Chemical Society 139}, 31 (Aug. 2017), 10909--10918.

\bibitem{Kraft}
{\sc Kraft, M.~A., Ohno, S., Zinkevich, T., Koerver, R., Culver, S.~P., Fuchs, T., Senyshyn, A., Indris, S., Morgan, B.~J., and Zeier, W.~G.}
\newblock Inducing high ionic conductivity in the lithium superionic argyrodites li6+xp1–xgexs5i for all-solid-state batteries.
\newblock {\em Journal of the American Chemical Society 140}, 47 (2018), 16330--16339.

\bibitem{Lejaeghere_SSSP}
{\sc Lejaeghere, K., Bihlmayer, G., Björkman, T., Blaha, P., Blügel, S., Blum, V., Caliste, D., Castelli, I.~E., Clark, S.~J., Corso, A.~D., de~Gironcoli, S., Deutsch, T., Dewhurst, J.~K., Marco, I.~D., Draxl, C., Dułak, M., Eriksson, O., Flores-Livas, J.~A., Garrity, K.~F., Genovese, L., Giannozzi, P., Giantomassi, M., Goedecker, S., Gonze, X., Grånäs, O., Gross, E. K.~U., Gulans, A., Gygi, F., Hamann, D.~R., Hasnip, P.~J., Holzwarth, N. A.~W., Iuşan, D., Jochym, D.~B., Jollet, F., Jones, D., Kresse, G., Koepernik, K., Küçükbenli, E., Kvashnin, Y.~O., Locht, I. L.~M., Lubeck, S., Marsman, M., Marzari, N., Nitzsche, U., Nordström, L., Ozaki, T., Paulatto, L., Pickard, C.~J., Poelmans, W., Probert, M. I.~J., Refson, K., Richter, M., Rignanese, G.-M., Saha, S., Scheffler, M., Schlipf, M., Schwarz, K., Sharma, S., Tavazza, F., Thunström, P., Tkatchenko, A., Torrent, M., Vanderbilt, D., van Setten, M.~J., Speybroeck, V.~V., Wills, J.~M., Yates, J.~R., Zhang, G.-X., and Cottenier, S.}
\newblock Reproducibility in density functional theory calculations of solids.
\newblock {\em Science 351}, 6280 (2016), aad3000.

\bibitem{Morgan}
{\sc Morgan, B.~J.}
\newblock Mechanistic origin of superionic lithium diffusion in anion-disordered li6ps5x argyrodites.
\newblock {\em Chemistry of Materials 33}, 6 (2021), 2004--2018.

\bibitem{Muy_2023}
{\sc Muy, S., Johnston, C., and Marzari, N.}
\newblock Aiida-defects: an automated and fully reproducible workflow for the complete characterization of defect chemistry in functional materials.
\newblock {\em Electronic Structure 5}, 2 (jun 2023), 024009.

\bibitem{Ohno_review}
{\sc Ohno, S., Banik, A., Dewald, G.~F., Kraft, M.~A., Krauskopf, T., Minafra, N., Till, P., Weiss, M., and Zeier, W.~G.}
\newblock Materials design of ionic conductors for solid state batteries.
\newblock {\em Progress in Energy 2}, 2 (mar 2020), 022001.

\bibitem{Ohno}
{\sc Ohno, S., Helm, B., Fuchs, T., Dewald, G., Kraft, M.~A., Culver, S.~P., Senyshyn, A., and Zeier, W.~G.}
\newblock Further evidence for energy landscape flattening in the superionic argyrodites li6+xp1–xmxs5i (m = si, ge, sn).
\newblock {\em Chemistry of Materials 31}, 13 (2019), 4936--4944.

\bibitem{ONG2013314}
{\sc Ong, S.~P., Richards, W.~D., Jain, A., Hautier, G., Kocher, M., Cholia, S., Gunter, D., Chevrier, V.~L., Persson, K.~A., and Ceder, G.}
\newblock Python materials genomics (pymatgen): A robust, open-source python library for materials analysis.
\newblock {\em Computational Materials Science 68\/} (2013), 314--319.

\bibitem{PBESol_Perdew}
{\sc Perdew, J.~P., Ruzsinszky, A., Csonka, G.~I., Vydrov, O.~A., Scuseria, G.~E., Constantin, L.~A., Zhou, X., and Burke, K.}
\newblock Erratum: Restoring the density-gradient expansion for exchange in solids and surfaces [phys. rev. lett. 100, 136406 (2008)].
\newblock {\em Phys. Rev. Lett. 102\/} (Jan 2009), 039902.

\bibitem{Prandini}
{\sc Prandini, G., Marrazzo, A., Castelli, I.~E., Mounet, N., and Marzari, N.}
\newblock {Precision and efficiency in solid-state pseudopotential calculations}.
\newblock {\em {NPJ COMPUTATIONAL MATERIALS} {4}\/} ({DEC 6} {2018}).

\bibitem{RayavarapuMAY}
{\sc Rayavarapu, P.~R., Sharma, N., Peterson, V.~K., and Adams, S.}
\newblock Variation in structure and li+-ion migration in argyrodite-type li6ps5x (x = cl, br, i) solid electrolytes.
\newblock {\em JOURNAL OF SOLID STATE ELECTROCHEMISTRY 16}, 5 (MAY), 1807--1813.

\bibitem{Sakuda}
{\sc Sakuda, A., Yamauchi, A., Yubuchi, S., Kitamura, N., Idemoto, Y., Hayashi, A., and Tatsumisago, M.}
\newblock Mechanochemically prepared li2s–p2s5–libh4 solid electrolytes with an argyrodite structure.
\newblock {\em ACS Omega 3}, 5 (2018), 5453--5458.

\bibitem{SCHLIPF201536}
{\sc Schlipf, M., and Gygi, F.}
\newblock Optimization algorithm for the generation of oncv pseudopotentials.
\newblock {\em Computer Physics Communications 196\/} (2015), 36--44.

\bibitem{SCHNEIDER2017151}
{\sc Schneider, H., Du, H., Kelley, T., Leitner, K., ter Maat, J., Scordilis-Kelley, C., Sanchez-Carrera, R., Kovalev, I., Mudalige, A., Kulisch, J., Safont-Sempere, M.~M., Hartmann, P., Wei\ss, T., Schneider, L., and Hinrichsen, B.}
\newblock A novel class of halogen-free, super-conductive lithium argyrodites: Synthesis and characterization.
\newblock {\em Journal of Power Sources 366\/} (2017), 151--160.

\bibitem{Stamminger}
{\sc Stamminger, A.~R., Ziebarth, B., Mrovec, M., Hammerschmidt, T., and Drautz, R.}
\newblock Ionic conductivity and its dependence on structural disorder in halogenated argyrodites li6ps5x (x = br, cl, i).
\newblock {\em Chemistry of Materials 31}, 21 (2019), 8673--8678.

\bibitem{TOPSAKAL2014263}
{\sc Topsakal, M., and Wentzcovitch, R.}
\newblock Accurate projected augmented wave (paw) datasets for rare-earth elements (re=la–lu).
\newblock {\em Computational Materials Science 95\/} (2014), 263--270.

\bibitem{VANSETTEN201839}
{\sc {van Setten}, M., Giantomassi, M., Bousquet, E., Verstraete, M., Hamann, D., Gonze, X., and Rignanese, G.-M.}
\newblock The pseudodojo: Training and grading a 85 element optimized norm-conserving pseudopotential table.
\newblock {\em Computer Physics Communications 226\/} (2018), 39--54.

\bibitem{DeePMD_kit}
{\sc Wang, H., Zhang, L., Han, J., and E, W.}
\newblock Deepmd-kit: A deep learning package for many-body potential energy representation and molecular dynamics.
\newblock {\em Computer Physics Communications 228\/} (2018), 178--184.

\bibitem{NC_pseudo}
{\sc Willand, A., Kvashnin, Y.~O., Genovese, L., V\'{a}zquez-Mayagoitia, A., Deb, A.~K., Sadeghi, A., Deutsch, T., and Goedecker, S.}
\newblock Norm-conserving pseudopotentials with chemical accuracy compared to all-electron calculations.
\newblock {\em The Journal of Chemical Physics 138}, 10 (2013), 104109.

\bibitem{YU2021105858}
{\sc Yu, C., Zhao, F., Luo, J., Zhang, L., and Sun, X.}
\newblock Recent development of lithium argyrodite solid-state electrolytes for solid-state batteries: Synthesis, structure, stability and dynamics.
\newblock {\em Nano Energy 83\/} (2021), 105858.

\bibitem{DeePMD_SE}
{\sc Zhang, L., Han, J., Wang, H., Saidi, W., Car, R., and E, W.}
\newblock End-to-end symmetry preserving inter-atomic potential energy model for finite and extended systems.
\newblock In {\em Advances in Neural Information Processing Systems\/} (2018), S.~Bengio, H.~Wallach, H.~Larochelle, K.~Grauman, N.~Cesa-Bianchi, and R.~Garnett, Eds., vol.~31, Curran Associates, Inc.

\bibitem{Northrup_formalism}
{\sc Zhang, S.~B., and Northrup, J.~E.}
\newblock Chemical potential dependence of defect formation energies in gaas: Application to ga self-diffusion.
\newblock {\em Phys. Rev. Lett. 67\/} (Oct 1991), 2339--2342.

\bibitem{SSB_review}
{\sc Zhang, Z., Shao, Y., Lotsch, B., Hu, Y.-S., Li, H., Janek, J., Nazar, L.~F., Nan, C.-W., Maier, J., Armand, M., and Chen, L.}
\newblock New horizons for inorganic solid state ion conductors.
\newblock {\em Energy Environ. Sci. 11\/} (2018), 1945--1976.

\bibitem{C8TA10790D}
{\sc Zhang, Z., Sun, Y., Duan, X., Peng, L., Jia, H., Zhang, Y., Shan, B., and Xie, J.}
\newblock Design and synthesis of room temperature stable li-argyrodite superionic conductors via cation doping.
\newblock {\em J. Mater. Chem. A 7\/} (2019), 2717--2722.

\bibitem{ZHENG2018198}
{\sc Zheng, F., Kotobuki, M., Song, S., Lai, M.~O., and Lu, L.}
\newblock Review on solid electrolytes for all-solid-state lithium-ion batteries.
\newblock {\em Journal of Power Sources 389\/} (2018), 198--213.

\bibitem{Zhou_Sb}
{\sc Zhou, L., Assoud, A., Zhang, Q., Wu, X., and Nazar, L.~F.}
\newblock New family of argyrodite thioantimonate lithium superionic conductors.
\newblock {\em Journal of the American Chemical Society 141}, 48 (2019), 19002--19013.

\end{thebibliography}

% \section{Supplementary Material}

\begin{titlepage}
    \begin{center}
        \vspace*{1cm}

        \LARGE
        \textbf{Supplementary materials}
        
        \vspace{1.0cm}
        \Huge
        Optimizing ionic conductivity of Lithium in Li$_7$PS$_6$ argyrodite via dopant engineering
            
       \vspace{1.5cm}

       % \textbf{Sokseiha Muy and Nicola Marzari}

       \vfill
       
       \vspace{0.8cm}

   \end{center}
\end{titlepage}

\beginsupplement

% \begin{figure}
% \centering
% \includegraphics[width=0.9\linewidth]{images/Divalent_dopant_300K.png}
% % \vspace{1cm}
%     \caption{Crystal structure of the cubic argyrodite of Li$_6$PS$_5$X with the halide X sites forming a closest packed host lattice in which PS$^{3-}_4$ tetrahedra occupy the octahedral voids and half of the tetrahedral voids are filled with the free S$^{2-}$. When X = Cl or Br, there is a partial
%     occupancy of the free S$^{2-}$ and X on the 4d and 4a sites which is believed to have significant influence on the long-range diffusion of Li ions \cite{Ohno_review}}. 
%     \label{fig:divalent_dopants}
% \end{figure}

\begin{figure}
\centering
\includegraphics[width=0.95\linewidth]{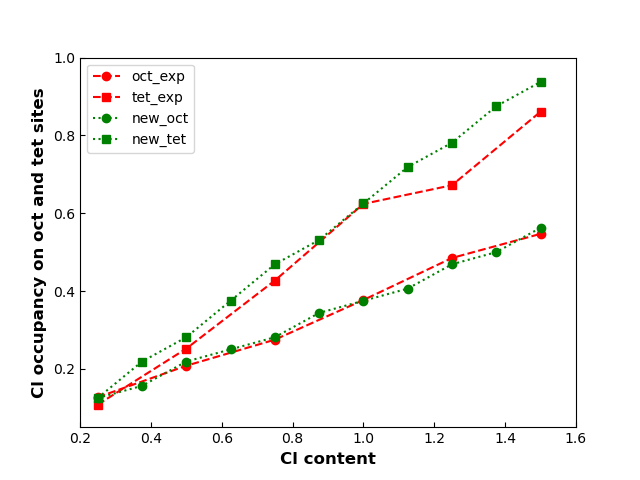}
    \caption{Comparison between the measured values of Cl occupancy on the octahedral and tetrahedral sites (green) \cite{gautam_Li7PS6}, and those used in the structural models in the MD simulations (red)}. 
    \label{fig:training_set_convergence}
\end{figure}

\end{document}